\documentstyle[wstwocl,epsf]{article}
\def\tmt{\times 10^{-2}}
\def\tmth{\times 10^{-3}}
\def\tmf{\times 10^{-4}}
\def\tmfv{\times 10^{-5}}
\newcommand{\beq}{\begin{equation}}
\newcommand{\eeq}{\end{equation}}
\newcommand{\bea}{\begin{eqnarray}}
\newcommand{\eea}{\end{eqnarray}}
\newcommand{\barr}{\begin{array}}
\newcommand{\earr}{\end{array}}
\newcommand{\bc}{\begin{center}}
\newcommand{\ec}{\end{center}}
\newcommand{\btab}{\begin{tabular}}
\newcommand{\etab}{\end{tabular}}
\newcommand{\gv}{\mbox{GeV}}
\newcommand{\tv}{\mbox{TeV}}
\newcommand{\mv}{\mbox{MeV}}

\newcommand{\nn}{\nonumber}
\newcommand{\ra}{\rightarrow}

\newcommand{\dro}{\Delta\rho}
\newcommand{\drqcd}{\delta\!\rho\,_{QCD}}
\newcommand{\roro}{\rho^{(2)}}

\newcommand{\al}{\alpha}

\newcommand{\G}{\Gamma}
\newcommand{\Gmu}{G_{\mu}}
\newcommand{\amu}{a_{\mu}}

\newcommand{\ganu}{\gamma_{\nu}}

\newcommand{\gafi}{\gamma_5}

\newcommand{\Pig}{\Pi^{\gamma}}

\newcommand{\noi}{\noindent}
\newcommand{\epmf}{e^+e^- \rightarrow f\bar{f}}

\newcommand{\epm}{e^+e^-}

\newcommand{\sm}{standard model }

\newcommand{\dal}{\Delta\alpha}

\newcommand{\mz}{M_Z^2}
\newcommand{\mw}{M_W^2}

\newcommand{\real}{\mbox{Re}}

\newcommand{\Dr}{\Delta r}

\newcommand{\eps}{\epsilon}

\newcommand{\aspi}{\frac{\alpha_s}{\pi}}

\newcommand{\alr}{A_{LR}}
\newcommand{\afb}{A_{FB}}

\newcommand{\ass}{asymmetries }

\newcommand{\pr}{Phys.\ Rev.\ }
 \newcommand{\prd}{Phys.\ Rev.\ D }
\newcommand{\zp}{Z.\ Phys.\ C }
\newcommand{\plb}{Phys.\ Lett.\ B }
 \newcommand{\prl}{Phys.\ Rev.\ Lett.\ }
\newcommand{\np}{Nucl.\ Phys.\ B }

\newcommand{\ms}{\overline{MS}}

\newcommand{\prs}{p_3^2}

\begin{document}
 
\bibliographystyle{unsrt}    % for BibTeX - sorted numerical labels
 
\def\CPbar{\hbox{{\rm CP}\hskip-1.80em{/}}}%temp replacement due to no font
 
\def\ap#1#2#3   {{\em Ann. Phys. (NY)} {\bf#1} (#2) #3.}
\def\apj#1#2#3  {{\em Astrophys. J.} {\bf#1} (#2) #3.}
\def\apjl#1#2#3 {{\em Astrophys. J. Lett.} {\bf#1} (#2) #3.}
\def\app#1#2#3  {{\em Acta. Phys. Pol.} {\bf#1} (#2) #3.}
\def\ar#1#2#3   {{\em Ann. Rev. Nucl. Part. Sci.} {\bf#1} (#2) #3.}
\def\cpc#1#2#3  {{\em Computer Phys. Comm.} {\bf#1} (#2) #3.}
\def\err#1#2#3  {{\it Erratum} {\bf#1} (#2) #3.}
\def\ib#1#2#3   {{\it ibid.} {\bf#1} (#2) #3.}
\def\jmp#1#2#3  {{\em J. Math. Phys.} {\bf#1} (#2) #3.}
\def\ijmp#1#2#3 {{\em Int. J. Mod. Phys.} {\bf#1} (#2) #3.}
\def\jetp#1#2#3 {{\em JETP Lett.} {\bf#1} (#2) #3.}
\def\jpg#1#2#3  {{\em J. Phys. G.} {\bf#1} (#2) #3.}
\def\mpl#1#2#3  {{\em Mod. Phys. Lett.} {\bf#1} (#2) #3.}
\def\nat#1#2#3  {{\em Nature (London)} {\bf#1} (#2) #3.}
\def\nc#1#2#3   {{\em Nuovo Cim.} {\bf#1} (#2) #3.}
\def\nim#1#2#3  {{\em Nucl. Instr. Meth.} {\bf#1} (#2) #3.}
\def\pcps#1#2#3 {{\em Proc. Cam. Phil. Soc.} {\bf#1} (#2) #3.}
\def\pl#1#2#3   {{\em Phys. Lett.} {\bf#1} (#2) #3.}
\def\prep#1#2#3 {{\em Phys. Rep.} {\bf#1} (#2) #3.}
\def\prev#1#2#3 {{\em Phys. Rev.} {\bf#1} (#2) #3.}
\def\prs#1#2#3  {{\em Proc. Roy. Soc.} {\bf#1} (#2) #3.}
\def\ptp#1#2#3  {{\em Prog. Th. Phys.} {\bf#1} (#2) #3.}
\def\ps#1#2#3   {{\em Physica Scripta} {\bf#1} (#2) #3.}
\def\rmp#1#2#3  {{\em Rev. Mod. Phys.} {\bf#1} (#2) #3.}
\def\rpp#1#2#3  {{\em Rep. Prog. Phys.} {\bf#1} (#2) #3.}
\def\sjnp#1#2#3 {{\em Sov. J. Nucl. Phys.} {\bf#1} (#2) #3.}
\def\spj#1#2#3  {{\em Sov. Phys. JEPT} {\bf#1} (#2) #3.}
\def\spu#1#2#3  {{\em Sov. Phys.-Usp.} {\bf#1} (#2) #3.}

\setcounter{secnumdepth}{2} % Number sections and subsections

%%%%%%%%%%%%%%%%%%%%%%%%%%%%%%%%%%%%%%%%%%%%%%%%%%
%                                                %
%    BEGINNING OF TEXT                           %
%                                                %
%%%%%%%%%%%%%%%%%%%%%%%%%%%%%%%%%%%%%%%%%%%%%%%%%%
 
\title{TESTS OF THE STANDARD MODEL}
 
\firstauthors{Wolfgang Hollik}
 
\firstaddress{Institut f\"ur Theoretische Physik, Universit\"at
Karlsruhe \\ D-76128 Karlsruhe, Germany}
 
%\secondauthors{ A.N. Other }
 
%if there are no second authors then comment out the line and adjust the
%maketitle command and \def\secondauthor... command in snow.sty
 
%\secondaddress{Department of Exotic Results, University of Heavens,
%Universe Road 999,\\  Whoknowswhere ZZZ123, Paradise}
 
%if there are no second authors then comment out the line and adjust the
%maketitle command and \def\secondaddress... command in snow.sty

\twocolumn[\maketitle\abstracts{
In this conference report a summary is given on the recent
theoretical work that has contributed to  improve the
theoretical predictions for testing the standard model in present
and future experiments. Precision calculations for the $Z$
resonance are reviewed and the status of the standard model is
discussed in the light of the recent top discovery and of the
results from precision experiments. Furthermore, theoretical
progress for the Higgs search, for $W$ physics at LEP II, and
for the anomalous magnetic moment of the muon is summarized.
New Physics beyond the minimal model is briefly discussed,
in particular the mimimal supersymmetric standard model in view of
the recent electroweak precision data. } 
                                             ]
 
\section{Introduction}
The present generation of high  precision experiments imposes
stringent tests on the standard model of electroweak and strong
interactions \cite{olshevsky}.
Besides the impressive achievements in the determination of the
$Z$ boson parameters \cite{olshevsky,lepewwg} and the $W$ mass
\cite{wmass}, the most important step has been the confirmation
of the top quark at the Tevatron \cite{top,d0,menzone}. Its mass
determination by the CDF collaboration \cite{top} yields
$m_t = 176 \pm 8 \pm 10$ GeV and by the D0 collaboration
 \cite{d0}:
$m_t = 199^{+19}_{-21} \pm 22$ GeV, resulting in an weighted average of
\beq            m_t = 180 \pm 12 \gv \, . \eeq
 
The high experimental
sensitivity in the electroweak observables, at the
level of quantum effects,  requires the highest standards
on the theoretical side as well. A sizeable amount of work has
contributed over the last few years to a steadily rising
improvement of the standard model predictions
pinning down the theoretical uncertainties to a level sufficiently
small for the current interpretation of the precision data, but
still sizeable enough to provoke conflict with a further increase
in the experimental accuracy.
 
\smallskip
In this report we review the recent theoretical work which has provided
the current theoretical basis for tests of the standard model at
present and future colliders.
A detailed description of the standard model predictions for
observables around the $Z$ peak
can be found in the Working Group
report on `Precision Calculations for the $Z$ Resonance'
\cite{yb95} (see also the talk by Bardin at this conference \cite{bardin}).
 This is followed by a discussion of the present status of the
standard model in the light of the recent experimental results.
Afterwards we report on theoretical progress in the study of
Higgs and $WW$ physics, and  for $g$-2 for muons. A short
discussion of the status of the MSSM concludes this presentation.

\section{Theory for precision tests}
\subsection{Radiative corrections}

The possibility of performing precision tests is based
on the formulation of the \sm as a renormalizable quantum field
theory preserving its predictive power beyond tree level
calculations. With the experimental accuracy in the investigation
of the fermion-gauge boson interactions being sensitive to the loop
induced quantum effects, also the more subtle parts of the \sm
Lagrangian are probed. The higher order terms
induce the sensitivity of electroweak observables
to the top and Higgs mass $m_t, M_H$
and to the strong coupling constant $\al_s$.

Before one can make predictions from the theory,
a set of independent parameters has to be determined from experiment.
All the practical schemes make use of the same physical input quantities
\beq \al, \; \Gmu,\; M_Z,\; m_f,\; M_H; \; \al_s \eeq   
for fixing the free parameters of the standard model.
 Differences between various schemes are formally
of higher order than the one under consideration.
 The study of the
scheme dependence of the perturbative results, after improvement by
resumming the leading terms, allows us to estimate the missing
higher order contributions.
 
\smallskip
Two fermion induced
large loop effects in electroweak observables deserve a special
discussion:
\begin{itemize}
\item
The light fermionic content of the subtracted photon vacuum polarization
$$
 \dal =   \Pig_{ferm}(0) -
     \real\,\Pig_{ferm}(\mz)
$$
corresponds to a QED induced shift
in the electromagnetic fine structure constant. The recent update of the
evaluation of the light quark content
by Eidelman, Jegerlehner \cite{eidelman} and Burkhardt, Pietrzyk
\cite{burkhardt} both yield the result
$$ (\dal)_{had} = 0.0280 \pm 0.0007 $$
and thus
confirm the previous value of \cite{vacpol} with an improved accuracy.
Other determinations \cite{swartz,martin}
agree within one standard deviation. Together with the leptonic
content, $\dal$ can
be resummed resulting in an effective fine structure
constant at the $Z$ mass scale:
\beq
   \al(\mz) \, =\, \frac{\al}{1-\dal}\,=\,
   \frac{1}{128.89\pm 0.09} \, .
\eeq
 \item
For a general structure of the scalar sector,
the electroweak mixing angle is related to the vector boson
masses  by
\bea
  \sin^2\theta & = &
 1-\frac{\mw}{\rho\mz}
   =
   1-\frac{\mw}{\mz} + \frac{\mw}{\mz} \dro \nn \\
 & \equiv & s_W^2 + c_W^2 \dro
\eea
where the $\rho$-parameter  $\rho = (1-\dro)^{-1}$
is an additional free parameter.
In the standard model,
one has the tree level relation $\rho=1$.
Loop effects, however, induce a deviation $\dro \neq 0$.
The main contribution is from the  $(t,b)$ doublet \cite{rho},
 in 1-loop and
 neglecting $m_b$ given by:
\beq
 \dro^{(1)} = 3 x_t, \;\;\;\; x_t =
 \frac{\Gmu m_t^2}{8\pi^2\sqrt{2}}
\eeq
Higher order
irreducible contributions have become available, modifying $\dro$
according to
 \beq
 \dro= 3 x_t \cdot [ 1+ x_t \,  \roro+ \drqcd ]
\eeq
 The electroweak 2-loop
 part \cite{bij,barbieri} is described by the
function $\roro(M_H/m_t)$ derived in \cite{barbieri} for general
Higgs masses.
$\drqcd$ is the QCD correction
to the leading $\Gmu m_t^2$ term
 \cite{djouadi,tarasov}
\beq
    \drqcd = -\,
\frac{\al_s(\mu)}{\pi}\, c_1
 +\left(\frac{\al_s(\mu)}{\pi}\right)^2 c_2(\mu)
\eeq
with
\beq
 c_1 =  \frac{2}{3} \left(
\frac{\pi^2}{3}+1\right)
\eeq
and the recently calculated
3-loop coefficent
 \cite{tarasov}
\beq
 c_2=-14.59 \;
 \mbox{  for } \mu =m_t \mbox{ and 6 flavors}
\eeq
with the on-shell top mass $m_t$.
It reduces the scale
dependence of $\drqcd$ significantly and hence is an important
entry to decrease the theoretical uncertainty of the standard model
predictions for precision observables (see section 2.4).
\end{itemize}
\subsection{The vector boson masses}
The correlation between
the masses $M_W,M_Z$ of the vector bosons          in terms
of the Fermi constant $\Gmu$, in 1-loop order given by
 \cite{sirmar}:
\beq
\frac{\Gmu}{\sqrt{2}}   =
            \frac{\pi\al}{2s_W^2 M_W^2} [
        1+ \Dr(\al,M_W,M_Z,M_H,m_t) ] \, .
\eeq
The decomposition
\beq
 \Dr = \Delta\al -\frac{c_W^2}{s_W^2}\,\dro^{(1)}
         + (\Dr)_{remainder} \, .
\eeq
separates the
leading fermionic contributions
                $\dal$ and $\dro$.
All other terms are collected in
the $(\Dr)_{remainder}$,
the typical size of which is of the order $\sim 0.01$.
 
\bigskip
The presence of large terms in $\Dr$ requires the consideration
of higher than 1-loop effects.
The modification of Eq.\ (10) according to
\bea
         1+\Dr & \, \ra\,& \frac{1}{(1-\Delta\al)\cdot
(1+\frac{c_W^2}{s_W^2}\dro) \, -\,(\Dr)_{remainder}}
 \nn \\
 & \equiv & \frac{1}{1-\Dr}
\eea
accommodates the following higher order terms
($\Dr$ in the denominator is an effective correction including
higher orders):
\begin{itemize}
\item
The leading log resummation \cite{marciano} of $\dal$:
$  1+\dal\, \ra \, (1-\dal)^{-1}$
\item
The resummation of the leading $m_t^2$ contribution \cite{chj}
in terms of $\dro$ in Eq.\ (6).
Beyond the $\Gmu m_t^2\al_s$ approximation through the $\rho$-parameter,
the complete
 $O(\al\al_s)$ corrections to the self energies
 are available from  perturbative  calculations
\cite{qcd} and by means of dispersion relations \cite{dispersion1}.
All the higher order terms contribute with the same positive sign
to $\Dr$,  thus
making the top mass dependence of $\Dr$ significantly flatter.
This is of high importance for the determination of an upper bound
on $m_t$ from $M_W$ measurements, which is affected by the order
of 10 GeV.
Quite recently, also non-leading terms to $\Delta r$ of
the type
$$ \Dr_{(bt)} = 3x_t \left(\frac{\al_s}{\pi}\right)^2 \left(
 a_1 \frac{\mz}{m_t^2} + a_2 \frac{M_Z^4}{m_t^4} \right)
$$
have been computed \cite{cks}. 
 For $m_t = 180$ GeV they contribute an extra
term of
$+0.0001$ to $\Dr$ and thus are within the uncertainty from
$\dal$.
\item
With the quantity $(\Dr)_{remainder}$ in the denominator
non-leading higher order terms
containing mass singularities of the type $\al^2\log(M_Z/m_f)$
from light fermions
are also incorporated \cite{nonleading}.
\end{itemize}

\subsection{$Z$ boson observables}
Measurements
of the $Z$ line shape in $\epmf$
yield the parameters
$M_Z,\, \G_Z$,   and the partial widths $\G_f$ or the peak
cross section
\beq
\sigma_0^f = \frac{12\pi}{\mz}\cdot\frac{\G_e\G_f}{\G_Z^2} \, .
\eeq
Whereas $M_Z$ is used as a precise input parameter, together
with $\al$ and $\Gmu$, the width, partial widths
and asymmetries allow
comparisons with the predictions of the standard model.
The predictions for the partial widths
as well as for the asymmetries
can conveniently be calculated in terms of effective neutral
current coupling constants for the various fermions.
 
\paragraph{\it Effective $Z$ boson couplings:}
 
The effective couplings follow
from the set of 1-loop diagrams
without virtual photons,
the non-QED  or weak  corrections.
These weak corrections
can conveniently be written
in terms of fermion-dependent overall normalizations
$\rho_f$ and effective mixing angles $s_f^2$
in the NC vertices \cite{formfactors}:
\bea
 & &
 J_{\nu}^{NC}  =   \left( \sqrt{2}\Gmu\mz \right)^{1/2} \,
  (g_V^f \,\ganu -  g_A^f \,\ganu\gafi)  \\
 & &
   =  \left( \sqrt{2}\Gmu\mz \rho_f \right)^{1/2}
\left( (I_3^f-2Q_fs_f^2)\ganu-I_3^f\ganu\gafi \right)  . \nn
\eea
%The complete expressions for $\rho_f,\kappa_f$ can be found in$^{27}$.
%Up to small terms negligible at the $Z$
%peak, they correspond to those of Bardin et al.$^{28}$.
$\rho_f$ and $s_f^2$ contain  universal
parts     (i.e.\ independent of the fermion species) and
non-universal parts which explicitly depend on the type of the
external fermions.
In their leading terms, incorporating also the next order,
the parameters are  given by
\beq
\rho_f  =  \frac{1}{1-\dro} + \cdots , \;\;\;
 s_f^2  = s_W^2 + c_W^2\,\dro + \cdots
\eeq
with $\dro$ from Eq.\ (6).
 
\smallskip
For the $b$ quark, also the non-universal parts have a strong
dependence on $m_t$ resulting from virtual top quarks in the
vertex corrections. The difference between the $d$ and $b$
couplings can be parametrized in the following way
\beq
  \rho_b = \rho_d (1+\tau)^2, \;\;\;\;
  s^2_b = s^2_d (1+\tau)^{-1}
\eeq
with the quantity
$$
 \tau = \Delta\tau^{(1)}
      + \Delta\tau^{(2)}
      + \Delta\tau^{(\al_s)}
$$
calculated perturbatively, at the present level comprising:
the complete 1-loop order term \cite{vertex} with $x_t$
from Eq.\ (5): 
\beq
\Delta\tau^{(1)} = -2 x_t - \frac{\Gmu\mz}{6\pi^2\sqrt{2}}
 (c_W^2+1)\log\frac{m_t}{M_W} + \cdots ,
\eeq
 the leading
electroweak 2-loop contribution of $O(\Gmu^2 m_t^4)$
\cite{barbieri,dhl}
\beq
\Delta\tau^{(2)} = -2\, x_t^2 \, \tau^{(2)} \, ,
\eeq
where
 $\tau^{(2)}$ is a function of $M_H/m_t$
with
 $\tau^{(2)} = 9-\pi^2/3$ for $M_H \ll m_t$;
the QCD corrections to the leading term of $O(\al_s\Gmu m_t^2)$
\cite{jeg}
\beq
\Delta\tau^{(\al_s)} =  2\, x_t \cdot \frac{\al_s}{\pi}
 \cdot \frac{\pi^2}{3} \, ,
\eeq
and the $O(\al_s)$ correction to the $\log m_t/M_W$ term in (17),
with a numerically very small coefficient \cite{log}.
 
\smallskip
\paragraph{\it Asymmetries and mixing angles:}
 
The effective mixing angles are of particular interest since
they determine the on-resonance asymmetries via the combinations
   \beq
    A_f = \frac{2g_V^f g_A^f}{(g_V^f)^2+(g_A^f)^2}  \, .
\eeq
Measurements of the \ass hence are measurements of
the ratios
\beq
  g_V^f/g_A^f = 1 - 2 Q_f s_f^2
\eeq
or the effective mixing angles, respectively.

\smallskip
\paragraph{\it $Z$ width and partial widths:}
 
The total
$Z$ width $\Gamma_Z$ can be calculated
essentially as the sum over the fermionic partial decay widths
\beq
 \Gamma_Z = \sum_f \, \Gamma_f + \cdots , \;\;\;\;
 \Gamma_f = \Gamma  (Z\ra f\bar{f})
\eeq
The dots indicate other decay channels which, however,
are not significant.
 The fermionic partial
widths,
 when
expressed in terms of the effective coupling constants
read up to 2nd order in the (light) fermion masses:
\bea
\Gamma_f
  & = & \G_0
 \, \left(
     (g_V^f)^2  +
     (g_A^f)^2 (1-\frac{6m_f^2}{\mz} )
                           \right)
 \cdot   (1+ Q_f^2\, \frac{3\al}{4\pi} ) \nn \\
     & +& \Delta\G^f_{QCD} \nn
\eea
with
$$
\G_0 \, =\,
  N_C^f\,\frac{\sqrt{2}\Gmu M_Z^3}{12\pi},
 \;\;\;\; N_C^f = 1
 \mbox{ (leptons)}, \;\; = 3 \mbox{ (quarks)}.
$$
The QCD correction for the light quarks
with $m_q\simeq 0$ is given by
\beq
 \Delta\G^f_{QCD}\, =\, \G_0
  \left( (g_V^f) ^2+ (g_A^f)^2 \right)
 \cdot K_{QCD}
\eeq
with \cite{qcdq}
$$
K_{QCD}  =   \frac{\al_s}{\pi} +1.41 \left(
  \frac{\al_s}{\pi}\right)^2 -12.8 \left(
  \frac{\al_s}{\pi}\right)^3
  -\frac{Q_f^2}{4}\frac{\al\al_s}{\pi^2} \,  .  
$$
For $b$ quarks
the QCD corrections are different due to  finite $b$ mass terms
and to top quark dependent 2-loop diagrams
 for the axial part:
\beq
 \Delta\G_{QCD}^b  =
 \Delta\G_{QCD}^d \, +\, \G_0 \left[
           (g_V^b)^2  \, R_V \,+\,
           (g_A^b)^2 \, R_A  \right]  \nn
\eeq
The coefficients in the perturbative expansions
\bea
 R_V &=& c_1^V \aspi + c_2^V (\aspi)^2 + c_3^V (\aspi)^3 + \cdots,
 \nn \\
 R_A &=&  c_1^A \aspi + c_2^A (\aspi)^2 + \cdots \,
 \nn
\eea
depending on $m_b$ and $m_t$,
are calculated up to third order
 in the vector and up to second order
in the axial part \cite{qcdb}.
 
Radiation of secondary fermions through photons
from the primary final state fermions
can yield another sizeable contribution to the partial $Z$ widths
which, however, is compensated by the corresponding virtual
contribution through the dressed photon propagator in the final state
vertex correction. For this compensation it is essential that
the analysis is inclusive enough, i.e.\ the cut to the invariant mass
of the secondary fermions is sufficiently large \cite{hoang}.
 
\subsection{Status of the standard model predictions}
 For a discussion of the theoretical reliability
of the \sm predictions one has to consider the various sources
contributing to their
uncertainties:

The experimental error propagating into the hadronic contribution
of $\al(\mz)$, Eq.\ (3), leads to
$\delta M_W = 13$ MeV in the $W$ mass prediction, and
$\delta\sin^2\theta = 0.00023$ common to all of the mixing
angles, which matches with the future experimental precision.

The uncertainties from the QCD contributions,
 besides the 3 MeV in the
hadronic $Z$ width, can essentially be traced back to
those in the top quark loops for the $\rho$-parameter.
They  can be combined into the following errors
\cite{kniehl95}, which have improved due to the recently available
3-loop result:
 
$$
 \delta(\dro) \simeq 1.5\cdot 10^{-4},   \;
 \delta s^2_{\ell} \simeq 0.0001
$$
for $m_t = 174$ GeV, and slightly larger for heavier top.
 
The size of unknown higher order contributions can be estimated
by different treatments of non-leading terms
of higher order in the implementation of radiative corrections in
electroweak observables (`options')
and by investigations of the scheme dependence.
Explicit comparisons between the results of 5 different computer codes  
based on  on-shell and $\ms$ calculations
for the $Z$ resonance observables are documented in the ``Electroweak
Working Group Report'' \cite{ewgr} in ref.\ \cite{yb95}
(see also the talk by Bardin \cite{bardin}).
The typical size of the genuine electroweak uncertainties
is of the order 0.1\%.
The following table  shows the uncertainty in a selected set of
precision observables. In particular for the very precise $s^2_e$
the theoretical uncertainty is still remarkable.
Improvements of the accuracy displayed in table 1
 require systematic electroweak and QCD-electroweak
 2-loop calculations.
As an example for the importance of electroweak non-leading
2-loop effects, an explicit calculation of these terms has been
performed for $\rho$ (the overall normalization) in neutrino
scattering \cite{padova}: they are sizeable and comparable to
the $O(\Gmu^2 m_t^4)$ term. Hence, one should take the registered       
uncertainties also for the $Z$ region very seriously.

\begin{table}[htbp]\centering
\caption[]
{Largest half-differences among central values $(\Delta_c)$ and among
maximal and minimal predictions $(\Delta_g)$ for $m_t = 175\,\gv$,
$60\,\gv < M_H < 1\,\tv$ and $\al_s(\mz) = 0.125$
(from ref.\ \cite{ewgr}) }
\vspace{0.5cm}
\begin{tabular}{c|c|c}
\hline \hline
Observable $O$ & $\Delta_c O$  & $\Delta_g O$ \\
\hline
            & & \\
$M_W\,$(GeV)          & $4.5\tmth$ & $1.6\tmt$\\
$\G_e\,$(MeV)          & $1.3\tmt$ & $3.1\tmt$\\
$\G_Z\,$(MeV)          & $0.2$     & $1.4$\\
$ s^2_e$             & $5.5\tmfv$ & $1.4\tmf$\\
$ s^2_b$             & $5.0\tmfv$ & $1.5\tmf$\\
$R_{had}$                 & $4.0\tmth$& $9.0\tmth$\\
$R_b$                 & $6.5\tmfv$ & $1.7\tmf$ \\
$R_c$                 & $2.0\tmfv$& $4.5\tmfv$ \\
$\sigma^{had}_0\,$(nb)    & $7.0\tmth$ & $8.5\tmth$\\
$\afb^l$             & $9.3\tmfv$ & $2.2\tmf$\\
$\afb^b$             & $3.0\tmf$ & $7.4\tmf$ \\
$\afb^c$             & $2.3\tmf$ & $5.7\tmf$ \\
$\alr$                & $4.2\tmf$ & $8.7\tmf$\\
\hline \hline
\end{tabular}
 
%\label{ta9}
\end{table}
%\normalsize
 
\smallskip \noi
Low angle
Bhabha scattering for a luminosity measurement at 0.1\% accuracy
still requires more theoretical effort. For a description of the present
status see the contributions by Jadach et al.\ and other authors
in \cite{yb95}. Presently an accuracy of 0.16\% has been claimed
\cite{bhlumi}.

\section{Standard model predictions versus data}
In table 2
the \sm predictions for $Z$ pole observables and the $W$ mass  are
put together. The first error corresponds to
the variation of $m_t$ in the observed range (1) and $ 60 < M_H < 1000$ GeV.
The second error is the hadronic
uncertainty from $\al_s=0.123\pm 0.006$, as measured
by QCD observables at the $Z$ \cite{alfas}.
 The recent combined LEP results \cite{olshevsky} on the $Z$ resonance
parameters, under the assumption of lepton universality,
are also shown in table 1, together with $s^2_e$ from
the left-right asymmetry at the SLC \cite{sld,olshevsky}.
 
The value for the leptonic mixing angle from the left-right asymmetry
$A_{LR}$ has come closer to the LEP result, but due to its smaller
error the deviation from the cumulative LEP average
is still $3\sigma$.

%\newpage
\begin{table*}[t]
            \caption{Precision observables: experimental results
%            \cite{olshevsky,lepewwg,wmass}
             (from refs.\ 1,2,3)
             and standard model         
             predictions. } \vspace{0.5cm}
            \bc
 \btab{|| l | l | r || }
\hline
\hline
 observable & exp. (1995) & \sm prediction \\
\hline
\hline
$M_Z$ (GeV) & $91.1884\pm0.0022$ &  input \\
\hline
$\Gamma_Z$ (GeV) & $2.4963\pm 0.0032$ & $2.4976 \pm 0.0077\pm 0.0033$ \\
%\hline
%$\Gamma_{had}$ (GeV) & $1.740\pm 0.008$ &
% $1.736\pm 0.008 \pm 0.007$ \\
%\hline
%$\Gamma_e$ (MeV) & $83.2\pm 0.4$ & $83.7\pm 0.4 $ \\
\hline
$\sigma_0^{had}$ (nb) & $41.4882\pm 0.078$ & $41.457\pm0.011\pm0.076$ \\
\hline
 $\G_{had}/\G_e$ & $20.788\pm 0.032 $ & $20.771\pm 0.019\pm 0.038$ \\
%\hline
%$\Gamma_e$ (MeV) & $83.82\pm 0.27$ & $83.7\pm 0.4 $ \\
\hline
$\Gamma_{inv}$ (MeV) & $499.9\pm 2.5$ & $501.6\pm 1.1$ \\
\hline
$\G_b/\G_{had}=R_b$  & $0.2219\pm 0.0017$ & $0.2155\pm 0.0004$ \\
\hline
$\G_c/\G_{had}=R_c$  & $0.1540\pm 0.0074$ & $0.1723\pm 0.0002$ \\
  \hline
$A_b$            & $0.841\pm 0.053$  & $0.9346 \pm 0.0006$ \\
\hline
$\rho_{\ell}$ & $1.0044\pm 0.0016$ & $1.0050\pm 0.0023$ \\
\hline
$s^2_{\ell}$ (LEP) & $0.23186\pm 0.00034$ & $0.2317\pm 0.0012$ \\
\hline
$s^2_e (A_{LR})$ & $0.23049\pm 0.00050$ & $0.2317\pm 0.0012$   \\
 LEP$+$SLC   &  $0.23143\pm 0.00028$    &                    \\
\hline
$M_W$ (GeV) & $80.26 \pm 0.16$ & $80.36\pm 0.18$  \\
\hline
\hline
\etab
\ec 
%  \vspace{-1.5cm}
\clearpage
\end{table*}

\smallskip \noindent
Significant deviations from the \sm predictions are observed in the
ratios  $R_b = \Gamma_b/\Gamma_{had}$ and
 $R_c = \Gamma_c/\Gamma_{had}$. The experimental values,
together with the top mass (1) from the Tevatron, are compatible
with the \sm at a confidence level of less than 1\% (see
\cite{olshevsky,behnke}),
enough to claim a deviation from the Standard Model.
The other precision observables are in perfect agreement with the
Standard Model. Note that
the experimental value for $\rho_{\ell}$ exhibits the presence of
genuine electroweak corrections by nearly 3 standard deviations.
 
The $W$ mass prediction is obtained  by Eqs.\ (10-12) from
 $M_Z,\Gmu,\al$ and  $M_H,m_t$.
  The quantity $s_W^2$ resp.\  the ratio $M_W/M_Z$
is indirectly measured in deep-inelastic neutrino scattering,
in particular in the
NC/CC neutrino         cross section ratio for isoscalar targets.
%The recent CCFR result \cite{bodek}
%  $$ s_W^2 =  0.2222 \pm 0.0057 $$
%combined with the CDHS and
%CHARM results \cite{neutrino} yields the world average \cite{bodek}
The present world average from CCFR, CDHS and CHARM,
including the new CCFR result
\cite{neutrino}
  $$ s_W^2 = 1-M_W^2/M_Z^2 = 0.2253 \pm 0.0047  $$
is fully consistent with the direct vector boson mass measurements
and with the standard theory.
 
\smallskip
        \paragraph{\it Standard model fits:}
 
Assuming the validity of the \sm a global fit to all electroweak
results from LEP, SLD, $p\bar{p}$ and $\nu N$
constrains the parameters $m_t,\al_s$ as follows:
\cite{olshevsky,lepewwg,busenitz}
\beq
    m_t = 178\pm 8^{+17}_{-20}\, \gv, \;\;\     
    \al_s = 0.123 \pm 0.004 \pm 0.002
\eeq
with $M_H= 300$ GeV for the central value.
The second error is from the variation of $M_H$
between 60 GeV and 1 TeV.
The fit results include the
uncertainties of the \sm calculations.
 
The indirect determination of the       
$W$ mass  from LEP/SLD data,
$$ M_W = 80.359\pm 0.055^{+0.013}_{-0.024} \, \gv \, , $$
is in best agreement with the direct measurement (see table 2). 
Moreover, the value obtained for
$\al_s$ at $M_Z$ coincides with the one
 measured from others than electroweak
observables at the $Z$ peak \cite{alfas}.

\medskip
The main Higgs dependence of the electroweak predictions is only
logarithmic in the Higgs mass. Hence, the sensitivity of the data
to $M_H$ is not very pronounced. Using the Tevatron value for $m_t$ as
an additional experimental constraint, the electroweak fit to all data
yields $M_H < 600$ GeV with approximately 95\% C.L.
\cite{olshevsky,lepewwg}.
Similar results with bounds on $M_H$ which are 100-200 MeV higher,
 based on the electroweak data from the winter
conferences,
 have been obtained in \cite{warsaw,jellis}.

\smallskip
\paragraph{\it Low energy results:}
The cross section for $\mu$-neutrino  electron scattering
and the electroweak mixing angle measured
by the CHARM II Collaboration  \cite{charm}
agree with the standard model values:
\bea
 \sigma(\nu e)/E_{\nu} &=& 16.51\pm 0.93)\cdot 10^{-42} cm^2 GeV^{-1}
 \nn \\
 (\mbox{SM} & : & 17.23\cdot 10^{-42} \, )     \nn \\
 \sin^2\! \theta_{\! \nu\! e} &=&  0.2324\pm 0.0083 \, .
\eea
The mixing angle  coincides with the result
on $s^2_{\ell}$ from the $Z$,
table 2, as expected by the theory. The major
loop contributions in the difference,
the different scales and the neutrino
charge radius, largely cancel each other by numerical coincidence
\cite{cradius}.
 
\smallskip
The recent results from the CLEO Collaboration   \cite{CLEO}
 on the flavor changing $B$ decays, the  branching ratio
 $BR(B\ra X_s \gamma) = (2.32\pm 0.67)\cdot 10^{-4} $ and the
inclusive photon spectrum, are fully consistent with the
\sm predictions based on the loop induced $b\ra s\gamma$ transition
\cite{bsgamma}.

\section{Future tests of the Standard Model}
\subsection{Higgs bosons}
The minimal model with a single scalar doublet is the simplest way
to implement the electroweak symmetry breaking. The experimental
result that the $\rho$-parameter is very close to unity is a
natural feature of models with doublets and singlets.
In the standard model, the mass $M_H$ of the Higgs boson
appears as the only additional parameter beyond the vector boson
and fermion masses. $M_H$ cannot be predicted but has to taken from
experiment. The present lower limit (95\% C.L.) from the search at
LEP \cite{grivaz} is 65 GeV.
Indirect determinations of $M_H$ from precision data have
already been discussed in section 3. The indirect mass bounds
depend sensitively on small changes in the input data, and their
reliability suffers at present from averaging data points which
fluctuate by several standard deviations. As a general feature,
it appears that the data prefer light Higgs bosons.
 
There are also a theoretical constraints on the Higgs mass
from vacuum stability and absence of a Landau pole \cite{lindner},
and from lattice calculations \cite{lattice}. A recent calculation
of the decay width for $H\ra W^+W^-$  in the large $M_H$ limit
in 2-loop order \cite{ghinculov} has shown that the 2-loop
contribution exceeds the 1-loop term in size (same sign) for
 $M_H > 930$ GeV. The requirement of applicability of
perturbation theory therefore puts a stringent upper limit on the
Higgs mass.
 
\smallskip
Higgs boson searches at LEP2 and future high energy hadron and
$\epm$ colliders require precise predictions for the Higgs
production and decay signatures together with detailed background
studies.   Improved calculations for the most relevant Higgs decays
 including higher order contributions have become available for
 
$\bullet$  $H \ra b\bar{b}$ for $M_H < 2m_t$ in order $O(\al_s^2)$
     \cite{larin,chetkwiat},
and $O(\al_s\Gmu m_t^2)$ \cite{hbb1} as well as
$O(\al_s^2\Gmu m_t^2)$ \cite{hbb2};
 
$\bullet$  $H\ra gg(g)$ up to order $O(\al_s^3)$ \cite{larin};
 
$\bullet$  $H\ra ZZ,WW$ for large $M_H$ in electroweak 2-loop order
 \cite{ghinculov}.
 
The QCD corrections for Higgs production at the LHC has been completed
\cite{lhc}, with the result of a significant enhancement by the
next order QCD contributions.
 
Higgs signal versus background studies were performed for the process
$\epm \ra b\bar{b}\ell\bar{\ell}$ at tree level order, both
(semi-)analytically \cite{leike} and by Monte Carlo methods
\cite{boos}. Current work is going on in the topical
LEP200 Workshop. An overview with more details on the present
theoretical status of Higgs physics can be found in these proceedings
\cite{haber}.
 
\subsection{$W$ bosons}
 
$W$ mass measurements at LEP2 with an error of about 40 MeV and tests
of the trilinear vector boson self couplings \cite{baur} require
standard model calculations for the process
$\epm \ra W^+W^- \ra 4 f$ and the corresponding 4-fermion background
processes at the accuracy level of 1\%.
A status report can be found in ref.\ \cite{fberends}.
One of the specific problems in the theoretical description of the
production process for off-shell $W$ bosons is the presence of the
width term in the $W$ propagator which violates gauge invariance,       
yielding gauge dependent amplitudes. As a solution it has been
proposed \cite{argyres} to take into account also the imaginary
part in the $WW\gamma$ vertex from the light fermion triangle
loops. This prescription is in accordance with gauge invariance
and cures the Ward identities between 2- and
3-point functions involving $W^{\pm}$ and $\gamma$.
 
\subsection{$g$-2 for muons}
The anomalous magnetic moment of the muon,
\beq
   a_{\mu} = \frac{g_{\mu}-2}{2}
\eeq
provides a precision test of the standard model at low energies.
Within the present experimental accuracy of
$\Delta\amu = 840\cdot 10^{-11}$, theory and experiment are in best
agreement, but the electroweak loop corrections are still hidden
in the noise. A new experiment, E 821 at Brookhaven National
Laboratory \cite{brookhaven}, is being prepared for 1996 to reduce
the experimental error down to $40\pm 10^{-11}$ and hence will
become sensitive to the electroweak loop contribution, which
at the 1-loop level \cite{ew1} amounts to $195\cdot 10^{-11}$.
 
For this reason the standard model prediction has to be known with
comparable precision. Recent theoretical work has contributed to
reduce the theoretical uncertainty by calculating the electroweak
2-loop terms \cite{ew21,ew22,ew23}
and updating the contribution from the hadronic
photonic vacuum polarization \cite{eidelman}
$$
 \amu^{had}(\mbox{vacuum pol.}) = (7024\pm 153)\cdot 10^{-11}
$$
which agrees within the error with the result of \cite{dub}.
The
main sources for the theoretical error at present are the hadronic
vacuum polarization and the light-by-light scattering mediated by
quarks, as part of the 3-loop hadronic contribution
\cite{sanda,bijnens}.
Table 3 shows  the breakdown of $\amu$. The hadronic part
is supplemented by the higher order $\al^3$
vacuum polarization effects \cite{had3} but is without the
light-by-light contribution, where the situation is unclear at
present.
 
\begin{table}[htbp]\centering
\caption[]
{Contributions $\Delta\amu$ to the muonic anomalous magnetic moment
and their theoretical uncertainties, in units of $10^{-11}$.  }
\vspace{0.5cm}
\begin{tabular}{l|r|r}
\hline \hline
source & $\Delta\amu$  & error \\
\hline
            & & \\
QED \cite{qed} & 116584708 & 5  \\
hadronic \cite{eidelman,had3}  & 6934  & 153 \\
EW, 1-loop \cite{ew1} & 195 &     \\
EW, 2-loop \cite{ew22} &   &    \\
(anomaly graphs) & -17   &  3 \\
EW, 2-loop \cite{ew21} &     &   \\
(all fermionic graphs) & -23 &  3  \\
EW, 2-loop \cite {ew23} &     &   \\
(bosonic graphs) & -17 & 2 \\
light-by-light \cite{sanda} & -36 &  16 \\
light-by-light \cite{bijnens} & -124 &  50 \\
\hline
future experiment &   & 40  \\
\hline \hline
\end{tabular}
 
%\label{ta9}
\end{table}
%\normalsize
 
The 2-loop electroweak contribution turns out to be as big in
size as the expected experimental error.
The dominating theoretical uncertainty at present is the error in
the hadronic vacuum polarization which can only be improved by new
measurements of the cross section for $\epm\ra hadrons$ in the low
energy range. But also the the contribution involving
light-by-light scattering needs clarification in order to reduce the
theoretical error.
 
\section{Precision data and ``New Physics''}
\paragraph{\it New physics in $R_b$?}
If the observed difference between the measured and calculated
values of $R_b$ is explained by a  non-standard contribution
$\Delta\G_b$ to the partial width $\G(Z\ra b\bar{b})$, then also
other hadronic quantities like $\G_Z, R_{had}, \dots$ are increased
unless the value of $\al_s$ is reduced simultaneously. Including
the new physics  $\Delta\G_b$ as an extra free parameter in the fit
yields the values \cite{olshevsky,lepewwg}:
 $$\al_s=0.102\pm 0.008, \;\;
 \Delta\G_b =11.7\pm3.8\pm1.4 \mv \, .$$
The top mass is affected only marginally, shifting the central
value by $+3$ GeV, but the impact on $\al_s$ is remarkable.

\paragraph{\it Virtual New Physics:}
The generalization of the aforementioned method consists in
the parametrization of the radiative corrections originating
 from the vector
boson self-energies in terms of the static $\rho$-parameter
$\dro(0) \equiv \eps_1$ 
and two other combinations of self-energies, $\eps_2$ and $\eps_3$
\cite{epsilon,abc1}. This
allows a more general analysis of the electroweak data
which accommodates extensions of the minimal model
affecting only the vector boson self-energies.
A further quantity $\eps_b$ has been introduced \cite{abc1} in order    
to parametrize specific non-universal
left handed contributions to the $Zbb$ vertex via
\beq
 g_A^b =g_A^d(1+\eps_b), \;\;\;
 g_V^b/g_A^b = (1-\frac{4}{3}s^2_d+\eps_b)\, (1+\eps_b)^{-1} \, .
\eeq
There is a wide literature \cite{pt} in this field  with various
conventions.
 
Phenomenologically, the $\eps_i$ are parameters which
can be determined experimentally from the electroweak precision data.
An updated analysis \cite{caravaglios}
on the basis of the recent electroweak results presented at this
conference \cite{olshevsky} yields for $\eps_b$ the value
$$
   \eps_b = 9.9 \pm 4.5 \;\;\; (\mbox{SM}:\;\; -6.6)    
$$
The large difference to the standard model value is an another way
of visualizing the deviation between the measured and predicted
number for the ratio $R_b$ (table 2).

\paragraph{\it The minimal supersymmetric standard model (MSSM):}
The MSSM deserves a special discussion
as the most predictive framework beyond the minimal model.
Its structure allows a similarly complete calculation of
the electroweak precision observables
as in the standard model in terms of one Higgs mass
(usually taken as $M_A$) and $\tan\beta= v_2/v_1$,
together with the set of
SUSY soft breaking parameters fixing the chargino/neutralino and
scalar fermion sectors.
It has been known since quite some time
\cite{higgs}
that light non-standard
Higgs bosons as well as light stop and charginos
% all around 50 GeV or little higher,
predict larger values for the ratio $R_b$ and thus diminish the
observed difference  \cite{susy1,susy3,susy4,susy5}.
Complete 1-loop calculations are meanwhile available for
$\Delta r$ \cite{susydelr} and for the $Z$ boson observables
\cite{susy3,susy4,susy5}.
 
The main results in view of the recent precision data are: \hfill
\\
$\bullet$ $R_c$ can hardly be moved towards the measured range.
 \hfill \\ 
$\bullet$ $R_b$ can come closer to the measured value, in particular
for light $\tilde{t}_R$ and light charginos. \hfill \\
 $\bullet$  $\al_s$ turns out to be smaller than in the 
 minimal model because of  the
reasons explained in the beginning of this section.
\hfill \\ 
$\bullet$ There are strong constraints from the other precision
observables which forbid parameter configurations shifting $R_b$
into the observed $1\sigma$ range.

For obtaining the optimized SUSY parameter set, therefore, a global
fit to all the electroweak precision data (including the top
mass measurements)
 has to be performed,
as done in refs.\ \cite{susy4,deboer}. Figure 1 displays the
experimental data normalized to the best fit results  
in the SM and MSSM, with 
the data from this conference \cite{deboer}.
For the SM, $\al_s$ identified with the experimental
number, therefore the corresponding result in Figure 1 is
centered at 1. The most relevant conclusions are: \hfill \\
(i) The difference between the experimental and theoretical value
of $R_b$ is diminished by a factor $\simeq 1/2$, \hfill \\
(ii) the central value for the strong coupling is $\al_s=0.110$
and thus is very close to the value obtained from deep inelastic
scattering, \hfill \\
(iii) the other observables are practically unchanged, \hfill \\
(iv) the $\chi^2$ of the fit is slightly better than in the minimal
model.

\setlength{\unitlength}{0.7mm}
\begin{figure}[hbt]
%\begin{picture}(100,110)(0,1)
%\mbox{\epsfxsize8.0cm\epsffile[0 20 595 794]{mssm.eps}}
\mbox{\epsfxsize8.0cm\epsffile{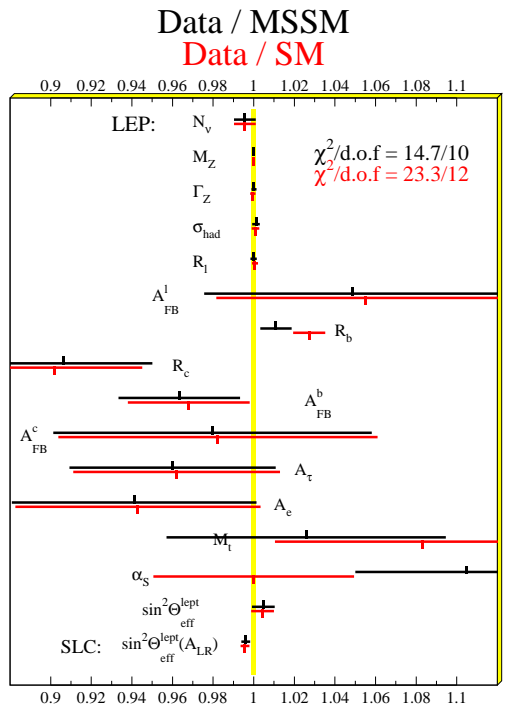}}
%\end{picture}
\caption{Experimental data normalized to the best fit results in
         the SM and MSSM }
\end{figure}

\section{Conclusions}
The experimental data for tests of the standard model 
have achieved an impressive accuracy.
In the meantime, many 
theoretical contributions have become available to improve and 
stabilize the \sm predictions. To reach, however, a theoretical
accuracy at the level 0.1\% or below, new experimental data on
$\dal$ and more complete electroweak 2-loop calculations are required.
The observed deviations of several $\sigma$'s in $R_b,R_c,\alr$
reduce the quality of the \sm fits significantly, but the
indirect  determination of $m_t$ is remarkably stable.
Still impressive is the perfect agreement between
theory and experiment for the whole set of the other
precision observables. SUSY can improve the situtation due to an 
enhancement of $R_b$ by new particles in the range of 100
GeV or even below, but it is not possible to accomodate $R_c$.
Within the MSSM analysis, the value for $\al_s$ is close to the one
from deep-inelastic scattering.   
In the QCD sector, a deviation from the theoretical expectation
in the inclusive jet cross section at the Tevatron has been reported
\cite{buckley} in terms of a significant excess of jets with
$E_T > 200$ GeV compared to the NLO QCD prediction. It is,
however, too early to draw conclusions about new physics from that.

\setcounter{secnumdepth}{0} %this ensures that there are no section numbers
                            %from here on in the text. Don't remove.
 
\section{Acknowledgments}
I wish to thank D. Bardin, W. de Boer, A. Czarnecki,
A. Dabelstein, R. Ehret, S. Meyer, A. Olshevsky, G. Passarino, D. Schaile 
for instructive  discussions and for support in preparing this review.

\end{document}